\documentclass[%
 reprint,
 nofootinbib,
 amsmath,amssymb,
 aps,
]{revtex4-2}

\usepackage{graphicx}
\usepackage{dcolumn}
\usepackage{bm}
\usepackage{booktabs}
\usepackage{multirow}
\usepackage{tikz}
\usepackage{tabularx}
\usepackage{placeins}
\usetikzlibrary{decorations.pathmorphing, decorations.markings, shapes.geometric}
\tikzset{
    boson/.style={decorate, decoration={snake, amplitude=0.5mm, segment length=1.5mm}},
    fermion/.style={decoration={markings, mark=at position 0.5 with {\arrow{latex}}}, postaction={decorate}},
    myblob/.style={ellipse, fill=gray!30, draw=black, minimum width=0.3cm, minimum height=0.7cm}
}

\newcommand{\wtopigamma}{W^{\pm} \to \pi^{\pm} \gamma}
\newcommand{\wtokgamma}{W^{\pm} \to K^{\pm} \gamma}
\newcommand{\wtorhogamma}{W^{\pm} \to \rho^{\pm} \gamma}

\newcommand{\ttbar}{t\bar{t}}

\newcommand{\gev}{\text{GeV}}
\newcommand{\tev}{\text{TeV}}

\newcommand{\pt}{p_\text{T}}
\newcommand{\dpia}{D_{\pi \gamma}}

\begin{document}

\preprint{APS/123-QED}

\title{``Hadron-in-fat-jet'' AI Tagging to Detect Rare Decays Such as $\wtopigamma$}

\author{Linrui Chen}
\email{chenlinrui@whu.edu.cn}
\affiliation{School of Physics, Peking University}%

\author{Tianyi Yang}
\email{tyyang99@pku.edu.cn}
\affiliation{School of Physics, Peking University}%

\author{Zixun Kou}
\affiliation{School of Physics, Peking University}%

\author{Zijian Wang}
\affiliation{School of Physics, Peking University}%

\author{Youpeng Wu}
\affiliation{School of Physics, Peking University}%

\author{Leyun Gao}
\affiliation{School of Physics, Peking University}%

\author{Qiang Li}
\email{qliphy0@pku.edu.cn}
\affiliation{School of Physics, Peking University}%

\date{\today}

\begin{abstract}
We investigate a novel class of boosted-object signatures at the LHC, where a high-$\pt$ fat-jet contains an identifiable hadron or quarkonium state originating from rare or semi-exclusive decays. Unlike conventional boosted jet studies, which focus on multi-prong partonic substructure, our approach probes hybrid configurations such as $\wtopigamma$, where a localized hadronic or quarkonium signal is embedded within a collimated jet. By fine-tuning the signature-oriented, pre-trained Sophon AI model optimized for large-radius jets, and combining it with an event-level BDT and a soft-drop-mass shape fit, we obtain an expected 95\% CL upper limit of ${\cal B}(\wtopigamma)<2.78\times10^{-5}$ for $450\,\mathrm{fb}^{-1}$ in our nominal setup. This study serves as a first proof-of-principle demonstration of the ``hadron-in-fat-jet'' paradigm; substantial gains in sensitivity are expected from improved trigger strategies, additional production channels, and dedicated taggers, while the methodology itself is broadly applicable to a wide range of rare Standard Model processes and searches for light or exotic resonances at present and future collider experiments.
\end{abstract}

\maketitle

\section{Introduction}

The Large Hadron Collider (LHC) provides a high-energy and high-luminosity environment for precision tests of the Standard Model (SM) and searches for rare processes. In the boosted regime, the decay products of heavy particles are collimated and can be reconstructed as single large-radius jets. Jet-substructure techniques have therefore become essential tools for identifying such objects, primarily targeting partonic decays such as $W/Z \to q\bar{q}$, $H \to b\bar{b}$, and top quark decays, where the relevant information is encoded in multi-prong radiation patterns~\cite{Almeida:2008yp,Marzani:2019hun,Larkoski:2017jix,Kogler:2018hem}.

However, much less attention has been paid to exclusive or semi-exclusive decay modes, in which identifiable hadrons emerge inside the jet and carry additional flavor-specific information beyond conventional prong-based observables. Exclusive hadronic decays of the $W$ boson constitute a particularly attractive example of this class, providing a unique laboratory for studying rare electroweak processes in the boosted regime. Their amplitudes involve both short-distance electroweak dynamics and long-distance QCD effects, making them sensitive probes of factorization, meson distribution amplitudes, and hadronic form factors at the electroweak scale~\cite{grossman_exclusive_2015,melia_exclusive_2016}. Phenomenological studies have also emphasized the potential of high-statistics LHC samples, in particular $t\bar{t}$ events, for probing rare exclusive $W$ decays~\cite{mangano_rare_2015}.

In the radiative decay $\wtopigamma$, the $W$ boson produces a light quark-antiquark pair that hadronizes into a charged pion, while a photon is radiated from the charged lines. The large $W$-boson mass makes this process a clean test bench for QCD factorization~\cite{grossman_exclusive_2015,PETERLEPAGE1979359}. In the Standard Model, the branching fraction is predicted to be as low as ${\cal O}(10^{-9})$~\cite{grossman_exclusive_2015}. When the $W$ boson is produced with large transverse momentum, the $\pi^{\pm}$ and $\gamma$ are emitted nearly collinearly, leading to a ``hadron-in-fat-jet'' topology with asymmetric energy sharing and non-perturbative internal structure.

Experimentally, existing searches have reconstructed $\wtopigamma$ as a resolved charged-hadron-plus-photon final state. CDF, CMS, and ATLAS have all performed dedicated searches~\cite{aaltonen_search_2012-1,sirunyan_search_2021,collaboration_search_2024}. The CMS search used $t\bar{t}$ events, where one $W$ boson provides a leptonic tag and the other is tested for the $\pi^{\pm}\gamma$ final state, and set ${\cal B}(\wtopigamma)<1.50\times10^{-5}$ at 95\% confidence level~\cite{sirunyan_search_2021}. The ATLAS search employed dedicated track-photon and diphoton triggers, obtaining ${\cal B}(\wtopigamma)<1.9\times 10^{-6}$, ${\cal B}(\wtokgamma)<1.7\times 10^{-6}$, and ${\cal B}(\wtorhogamma)<5.2\times10^{-6}$ at 95\% confidence level~\cite{collaboration_search_2024}.

While these resolved analyses have established the current experimental bounds, boosted topologies provide a complementary search regime. When the $W$ boson carries a sufficiently large transverse momentum, the pion and photon become highly collimated and may be reconstructed within a single large-radius jet. In this case, jet-substructure techniques offer an alternative approach that exploits the internal structure of the merged jet, potentially extending the reach of rare hadronic decay searches and enabling sensitivity to a broader class of ``hadron-in-fat-jet'' signatures.

In this work, we study $\wtopigamma$ in $W+$jets production, where the large inclusive production rate, combined with recoil against QCD radiation, provides access to boosted $W$ bosons. We treat the merged $\pi^{\pm}\gamma$ system as a large-radius jet with radius parameter $R=0.8$, and use jet-level information to identify the rare semi-exclusive decay. Specifically, we use $\wtopigamma$ as a benchmark to explore the fine-tuning of the existing Sophon framework for ``hadron-in-fat-jet'' signatures, and to provide a proof-of-principle demonstration of ``hadron-in-fat-jet'' tagging for the study of rare decay processes. We also provide a first phenomenological sensitivity estimate for boosted $\wtopigamma$ based on realistic jet-level fast simulation. This boosted channel could be combined with existing resolved searches in future experimental analyses.

\begin{figure}
    \centering

\begin{tikzpicture}[scale=1.2]
        \coordinate (i) at (0,0);         
        \coordinate (v1) at (1.2,0);      
        \coordinate (v2) at (2.4,0.8);    
        \coordinate (f1) at (3.4, 1.4);   
        \node[myblob] (b) at (3.8,0) {};  
        \coordinate (f2) at (5.0,0);      

        \draw[boson] (i) -- node[above] {$W^+$} (v1);
        \draw[fermion] (v1) -- node[above] {$u$} (v2); 
        \draw[fermion] (v2) -- (b);                     
        \draw[boson] (v2) -- node[above] {$\gamma$} (f1); 

        \draw[fermion] (b) to[out=-135, in=-45] node[below] {$\bar{d}$} (v1);

        \draw[double, double distance=1pt] (b) -- node[above] {$\pi^+$} (f2);
    \end{tikzpicture}

    \hfill

        \centering
\begin{tikzpicture}[scale=1.2]
                \coordinate (i) at (0,0);         
        \coordinate (v1) at (1.0,0);      
        \coordinate (f1) at (1.8,-1.0);   
        \coordinate (v2) at (2.4,0);      
        \node[myblob] (b) at (3.8,0) {};  
        \coordinate (f2) at (5.0,0);      

        \draw[boson] (i) -- node[above] {$W^+$} (v1);
        \draw[boson] (v1) -- (v2);        
        \draw[boson] (v1) -- node[right] {$\gamma$} (f1);

        \draw[fermion] (v2) to[out=45, in=135] node[above] {$u$} (b);
        \draw[fermion] (b) to[out=-135, in=-45] node[below] {$\bar{d}$} (v2);

        \draw[double, double distance=1pt] (b) -- node[above] {$\pi^+$} (f2);
        \end{tikzpicture}
        \caption{Leading-order Feynman diagrams for the radiative exclusive decay $\wtopigamma$. }
\end{figure}
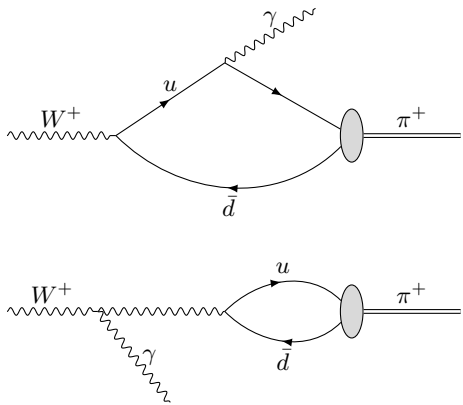

\section{Experimental setup}

Our study is based on simulated datasets for the LHC $pp$ collisions at center-of-mass $\sqrt{s}=13\,\tev$. We defined the signal phase space by selecting events featured by a high-$\pt$ fat-jet with a large electromagnetic energy fraction, optimized for capturing the boosted $\wtopigamma$ signature. Our simulated datasets contained the boosted $\wtopigamma$ signal from $W+$jets and various backgrounds including $\ttbar$, $W\to \tau \nu$, $W+$jets with hadronic and leptonic $W$ decays, $Z+$jets with $Z$ to $l^+l^-$, $\nu\bar{\nu}$, $q\bar{q}$, $W+\gamma$ and QCD multijets. Hard-scattering events were first generated at parton-level with \textsc{MadGraph5\_aMC@NLO}~\cite{alwall2014automated}, which provides matrix-element calculations and parton-level kinematics for both signal and background processes. To accurately model the proton structure, we employed the NNPDF 3.1 next-to-next-to-leading-order (NNLO) parton distribution function (PDF) set~\cite{ball2017parton}. The generated events were then interfaced to \textsc{Pythia8} for parton showering, hadronization and particle decays, modeling the transition from a few-body final state to a realistic multi-hadron configuration~\cite{Sjostrand:2007gs}.

For the fast detector simulation, we employed a custom \textsc{Delphes} 3.5.1~\cite{deFavereau:2013fsa} configuration explicitly optimized for the boosted topology and machine learning applications, following the \textsc{JetClass-II} framework~\cite{li_accelerating_2024}. Standard isolated object identification and flavor-tagging algorithms were bypassed. Instead, the simulation incorporates an average of 50 pileup interactions and utilizes the Pileup Per Particle Identification (PUPPI) algorithm~\cite{bertolini2014pileup} to mitigate pileup contamination. The track resolutions are realistically smeared to match the Phase-I CMS detector conditions. The PUPPI-weighted energy-flow objects are then clustered into large-radius ($R=0.8$) fat-jets using the anti-$k_T$ algorithm named as ``AK8-jet'' with the minimum $\pt$ thresholds of $200\, \gev$, while jets with low-radius ($R=0.4$) are also maintained as ``AK4-jet''. Crucially, the low-level kinematic and identity features of constituent particles within these fat-jets are retained, providing the essential granular representation required for the jet-tagging model.

Table~\ref{tab:mcsample} details the number of generated events and the corresponding production cross-sections used for the signal and background normalization. To enrich the boosted phase space and emulate the high-$H_\text{T}$ trigger region, we impose a common reconstruction-level skimming requirement, $H_\text{T}^{\rm reco}>1050\,\gev$, where $H_\text{T}^{\rm reco}$ is defined as the scalar sum of the transverse momenta of reconstructed jets~\cite{hayrapetyan2024performance}. This requirement is applied to all samples used in the analysis, including QCD multijets. For the signal and the non-QCD background samples, the selection is applied as an offline skim after event generation and detector simulation. For the QCD multijet sample, whose inclusive rate is much larger, the same requirement is applied during the reconstruction/skimming workflow so that only events passing the requirement are retained. The finite Monte Carlo statistics, especially for the QCD sample are accounted for by the background-yield smoothing and shape-extrapolation procedure described in Sec.~V.

\begin{table}[!h]
        \centering
        \caption{Summary of generated MC samples, detailing the decay processes, number of generated events ($N_{\text{gen}}$), and their corresponding cross-sections.}
        \vspace{0.5em}
        \begin{tabular*}{\linewidth}
        {@{\extracolsep{\fill}}ccc}
            \toprule
            \textbf{Decay Processes} & \textbf{$N_{\text{gen}}(\times 10^5)$} & \textbf{Cross-Section (mb)}\\
            \midrule
            Signal ($W^{\pm}\to \pi^{\pm}\gamma$) & 99 & $1.632\times 10^{-8}$ \\
            \midrule
            QCD & 9902.5 & $9.047\times 10^{-6}$\\
            $W^{\pm}\to \tau^{\pm} \nu$ & 20 & $1.629\times 10^{-8}$\\
            $p\ p \to t \bar{t}$ & 201 & $1.523\times 10^{-8}$\\
            $p\ p \to W \gamma$ (ISR \& FSR) & 1 & $2.587 \times 10^{-11}$\\
            $p\ p \to W[\to l\nu]+\text{jets}$ & 40 & $4.576\times 10^{-9}$ \\
            $p\ p \to W[\to q\bar{q}]+ \text{jets}$ & 301 & $3.461\times 10^{-8}$\\
            $p\ p \to Z[\to l^+l^-]+\text{jets}$ & 5 & $4.922\times 10^{-10}$\\
            $p\ p \to Z[\to \nu\bar{\nu}]+\text{jets}$ & 10 & $1.189\times 10^{-9}$\\
            $p\ p \to Z[\to q\bar{q}]+\text{jets}$ & 100 & $1.407\times 10^{-8}$\\
            \bottomrule
        \end{tabular*}
        \label{tab:mcsample}
\end{table}

\section{Sophon model Fine-tuning}

The Sophon framework (Signature-Oriented Pre-training for Heavy-resonance ObservatioN) introduces a new paradigm for resonance searches at the LHC~\cite{li_accelerating_2024,zhao_novel_2025}. Conventional jet taggers are typically optimized for specific signal hypotheses, limiting their applicability to a narrow class of final states~\cite{cms2020identification,qu2020jet}. In contrast, Sophon employs a signature-oriented strategy, in which a deep neural network is pre-trained on a large and diverse set of jet topologies to learn a universal representation of jet substructure.

The model is trained on a high-dimensional classification task spanning 188 jet categories, enabling it to encode both perturbative radiation patterns and non-perturbative features within a common latent space. This representation can be reused across analyses via transfer learning or by constructing discriminants directly from the model outputs, thereby providing a flexible and model-agnostic search framework.

This approach is particularly well suited to non-standard jet topologies, such as the ``hadron-in-fat-jet'' configurations considered in this work, where an identified hadron or quarkonium is embedded within a boosted fat-jet. Such signatures involve asymmetric energy sharing and non-perturbative dynamics that are not fully captured by traditional prong-based observables. In this context, the generalized feature space learned by Sophon offers a promising avenue to enhance sensitivity and systematically explore these novel signatures.

To identify the ``hadron-in-fat-jet'' topology, we fine-tune the original Sophon model, which initially featured ${\cal O}(100)$ categories but lacked specific hadronic final-state nodes. We explicitly incorporate 18 new categories (yielding a total of 206 categories for the fine-tuned model). These newly added nodes include signal and background final states that mimic the boosted $\wtopigamma$ signature, such as $\tau_h\nu$\footnote{Here $\tau_h\nu$ denotes an auxiliary AK8 fat-jet category used in the fine-tuning label space, rather than a standard narrow-cone hadronic-$\tau$ identification category such as those used in HPS/DeepTau-style reconstruction. Ordinary boosted $\tau_h\nu$ configurations are largely suppressed by the soft-drop-mass requirement; this auxiliary node is included to penalize residual $\tau$-like topologies that can leak into the $\pi\gamma$ signal-like region.}, $\pi\gamma q$ and $\pi\gamma \tau_h \nu$, along with other topologies susceptible to being misidentified as the signal. The training dataset was generated from the decays of a hypothetical spin-0 resonance, $X^{0, \pm}$, with its mass $m_X$ uniformly sampled between $15$ and $500\,\gev$. This broad and uniform mass sampling is deliberately implemented to prevent the Sophon model from learning kinematic features correlated with the resonance mass. Consequently, it forces the model to strictly rely on the internal multi-prong substructure and energy distribution of the fat-jet for classification, ensuring a mass-decorrelated tagging performance.

The fine-tuning procedure is executed in two stages. First, the network backbone is frozen, while only the final classification head is updated. This stage employs a relatively large learning rate over a few epochs to warm up the weights of the newly introduced categories, ensuring compatibility with the pre-trained weights. Subsequently, the entire network is unfrozen and trained with a reduced learning rate over a longer duration. This enables the model to adapt to the new categories while preserving its discriminative power for the original ones.


\section{Analysis strategies}

To isolate this rare signal from overwhelming Standard Model backgrounds, we employ a multi-step selection strategy focused on both jet-level properties and event-level kinematics, paired with a robust background model. First, candidate large-radius jets are required to fall within a kinematic window of $\pt>200\, \gev$ and $|\eta|<2.5$. To ensure the jet substructure is consistent with a $W$ boson decay, a minimum soft-drop-mass ~\cite{dasgupta2013towards,larkoski2014soft} requirement of $50\,\gev$ is imposed. This threshold effectively suppresses low-mass QCD jets while retaining the majority of the $W$ boson signal.

The core of our discrimination power resides in a fine-tuned Sophon model. Leveraging its signature-oriented architecture, we construct a dedicated, highly optimized discriminant, $\dpia$, to maximize the separation between the $\pi\gamma$ signal and various background categories. The discriminant is formulated as
\begin{equation}
    \dpia = \frac{g_{X\to \pi\gamma}}{g_{X\to \pi \gamma} + g_{X\to \pi \gamma q} + g_{X\to \tau_h \nu} + g_{QCD} + \sum_{P_1}g_{P_1}}
\end{equation}
where $g_i$ represents the output scores from the 206-category fine-tuned Sophon model. In this optimized formulation, the numerator is strictly driven by the pure $X\to \pi\gamma$ category to ensure high signal purity. The denominator acts as a targeted background penalty, comprising the signal node itself alongside primary mimicking topologies: the $X\to \pi\gamma q$ decays, hadronic tau decays $\tau_h \nu$, QCD multijets, and a collectively optimized subset of background categories denoted as $P_1$.

The composition of the $P_1$ subset was systematically determined through a data-driven evaluation of background leakage in the high-$\dpia$-score region. In an earlier baseline iteration, the discriminant's penalty term was limited to the signal-like nodes, QCD multijets, and hadronic tau ($\tau_h \nu$) categories. While this baseline effectively suppressed the multijet and $\tau_h \nu$ topologies, rigorous sensitivity scans---specifically evaluating background events that survived a stringent threshold of $D_{\pi\gamma}^{\text{baseline}} > 0.9$---revealed significant residual contamination from $Z+$jets, $W+$jets, and $t\bar{t}$ processes in the signal region.

To understand and mitigate this leakage, we analyzed the fine-tuned Sophon model's multi-class output distributions for these surviving background events. We observed that they spuriously peaked in specific non-target categories. For instance, leptonic $Z+$jets decays predominantly populated pure electron- or muon-like 2-prong nodes (e.g., $X \to ee, \mu\mu$), whereas $t\bar{t}$ and $W+$jets events artificially populated complex hybrid nodes, such as $X\to \pi\gamma  qq$ or heavy-flavor configurations. Consequently, we explicitly grouped these highly populated \lq fake\rq \ nodes into the $P_1$ subset and introduced them into the denominator of the refined discriminant. This targeted penalty actively and efficiently eliminated the dominant Standard Model contaminants most prone to mimicking the boosted $\wtopigamma$ signature.
The detailed composition of the strategically targeted $P_1$ subset is summarized in Table ~\ref{tab:p1}.

\begin{table}[htbp]
    \centering
    \caption{Summary of the first-priority background categories ($P_1$) incorporated into the denominator of the optimized $\dpia$ discriminant. These specific 2-prong and 3- or 4-prong decay topologies were systematically identified by analyzing the surviving $Z+$jets, $W+$jets and $t\bar{t}$ background events that bypassed initial baseline selections. Explicitly penalizing these nodes efficiently eliminates the dominant standard model configurations most prone to mimicking the boosted $\wtopigamma$ signature.}
    \begin{tabular*}{\linewidth}{@{\extracolsep{\fill}}llll}
        \toprule
        \multicolumn{4}{c}{First Priority labels $P_1$}\\
        \midrule
        $X\to$ 2-prong & $X\to ee$ & $X\to \mu\mu$ &  \\
        \midrule
        \multirow{6}{*}{$X\to3$- or $4$-prong} & $X\to \pi\gamma qq$ & $X\to \pi\gamma qs$ & $X\to \pi\gamma cq$ \\
         & $X\to \pi\gamma cs$ & $X\to qqee$ & $X\to qq\mu\mu$ \\
         & $X\to see$ & $X\to qee$ & $X\to gee$ \\
         & $X\to q\mu\mu$ & $X\to bqe\nu$ & $X\to sqe\nu$ \\
         & $X\to qqe\nu$ & $X\to bq\mu\nu$ & $X\to sq\mu\nu$ \\
         & $X\to qq\mu\nu$ & & \\
        \bottomrule
    \end{tabular*}
    \label{tab:p1}
\end{table}

\begin{figure}
    \centering
    \includegraphics[width=0.95\linewidth]{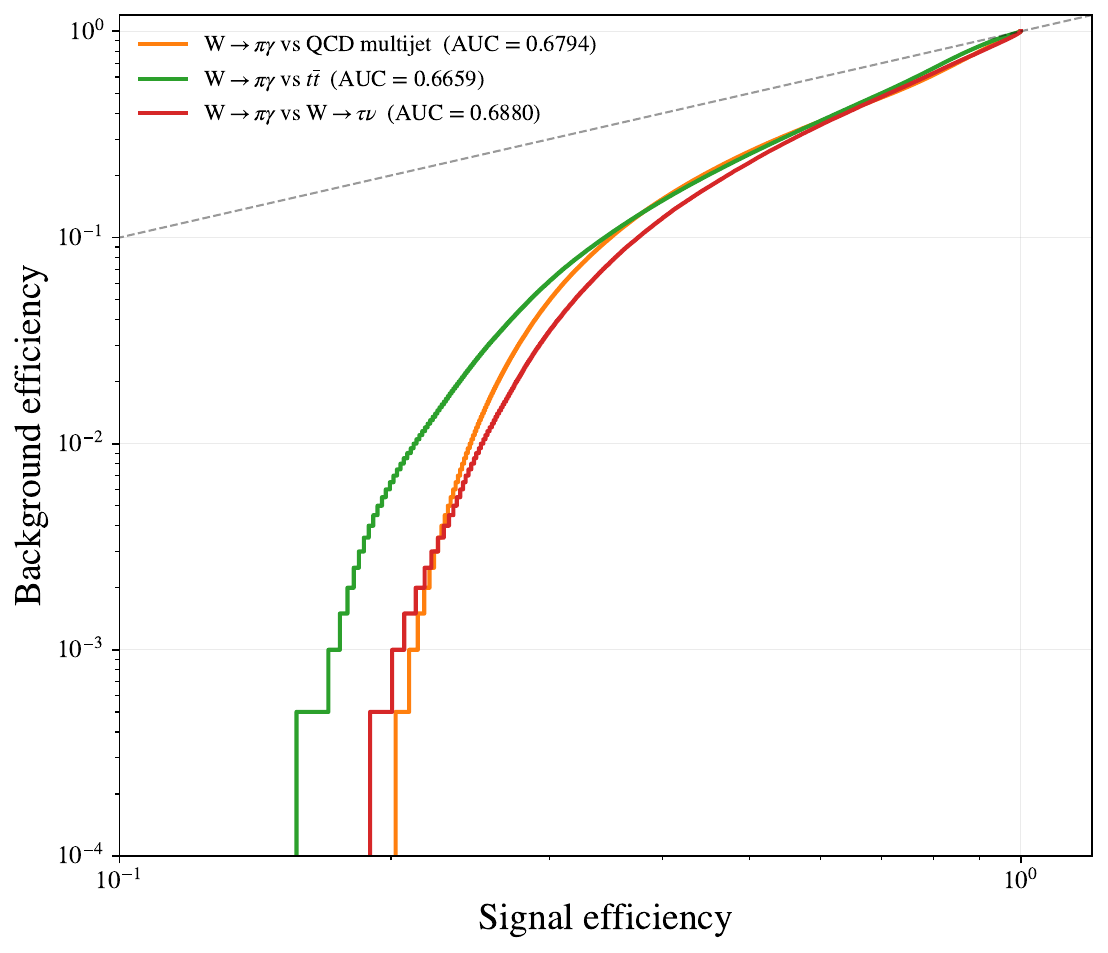}
    \caption{ROC curves evaluating the tagging performance of the $\dpia$ discriminant. The plot illustrates the signal efficiency for $\wtopigamma$ against the background efficiencies of QCD multijets, $t\bar{t}$ and $W\to \tau\nu$.}
    \label{fig:ROC}
\end{figure}

To evaluate the efficacy of the Sophon model fine-tuning and the physical soundness of the $\dpia$ form, Receiver Operating Characteristic (ROC) curves were generated for the three primary background processes that most heavily mimic the $\wtopigamma$ signal---namely QCD, $t\bar{t}$, and $W\to \tau\nu$---as depicted in Fig.~\ref{fig:ROC}. The corresponding Area Under the Curve (AUC) metrics indicate a robust and stable background rejection capability across diverse fake-signal topologies, further validating the effectiveness of the targeted penalty terms incorporated into the $\dpia$ formulation.

While the fine-tuned Sophon model provides exceptional discrimination power at the fat-jet level, the ultimate sensitivity of the analysis can be further enhanced by exploiting event-wide correlations. To this end, an event-level Boosted Decision Tree (BDT) classifier is implemented using the XGBoost framework ~\cite{chen2016xgboost}. The BDT is trained to separate the $\wtopigamma$ signal from an inclusive mixture of all standard model backgrounds.

The input features for the BDT comprise macroscopic event kinematics, object multiplicities, and angular correlations, whose variable names and physical meanings are listed in Table ~\ref{tab:inputvar}. Crucially, to facilitate reliable data-driven background estimation and in-situ calibration, these input variables are explicitly chosen to be orthogonal (i.e., highly uncorrelated) to the jet-level Sophon discriminant $\dpia$. This orthogonality condition ensures that stringent selections on the BDT score will not artificially sculpt the $\dpia$ background distributions, preserving the viability of control regions.

\begin{table}[htbp]
    \centering
    \footnotesize
    \caption{Summary of the input variables used for the event-level BDT classifier and their physical meanings.}
    \begin{tabular*}{\linewidth}{@{\extracolsep{\fill}}cc}
        \toprule
        \textbf{Variables} & \textbf{Physical Meaning}\\
        \midrule
        $p_{T,\,\text{AK4}}^{\text{lead}}$ & $\pt$ of the leading AK4-jet outside the fat-jet\\
        $H_{\text{T,$\,$AK4}}$ & $H_{\text{T}}$ of all AK4-jets outside the fat-jet\\
        $|\eta_l|$ & Absolute value of the lepton pseudorapidity\\
        $I_{\mu}^{\text{tight}}$ & Selection flag indicating the presence of tight $\mu$\\
        $I_e^{\text{tight}}$ & Selection flag indicating the presence of tight $e$\\
        $I_l^{\text{rel}}$ & Relative isolation of the lepton\\
        $\Delta R_{\text{min}}$ & $\Delta R$ between any AK4-jet and the fat-jet\\
        $p_{\text{T,}\,\text{lead}}^{\gamma,\,\text{lead}}$ & $\pt$ of the leading photon inside the leading fat-jet\\
        $\Delta \phi(l,\,\text{FJ})$ & $\Delta \phi$ between the lepton and the fat-jet\\
        $N_{\text{AK4}}$ & Multiplicity of AK4-jets outside the fat-jet\\
        $p_{\text{T}}^{\text{miss}}$ & Magnitude of the missing $\pt$\\
        $\Delta \phi (p_{\text{T}}^{\text{miss}},\,\text{FJ})$ & $\Delta \phi$ between the $p_{\text{T}}^{\text{miss}}$ and the fat-jet\\
        \bottomrule
    \end{tabular*}
    \label{tab:inputvar}
\end{table}

Special attention is given to the event weighting during the BDT training process. Initially, both signal and background events are scaled using physical weights, corresponding to an expected integrated luminosity of $450$ fb$^{-1}$. To prevent the overwhelmingly large background yields from dominating the gradient boosting process, a dedicated training weight is applied, normalizing the total sum of weights for both the signal and the inclusive background to unity. Conversely, the model evaluation on the testing dataset strictly utilizes the original physical weights to reflect realistic experimental conditions.

\begin{figure}[htbp]
    \centering
    \includegraphics[width=0.95\linewidth]{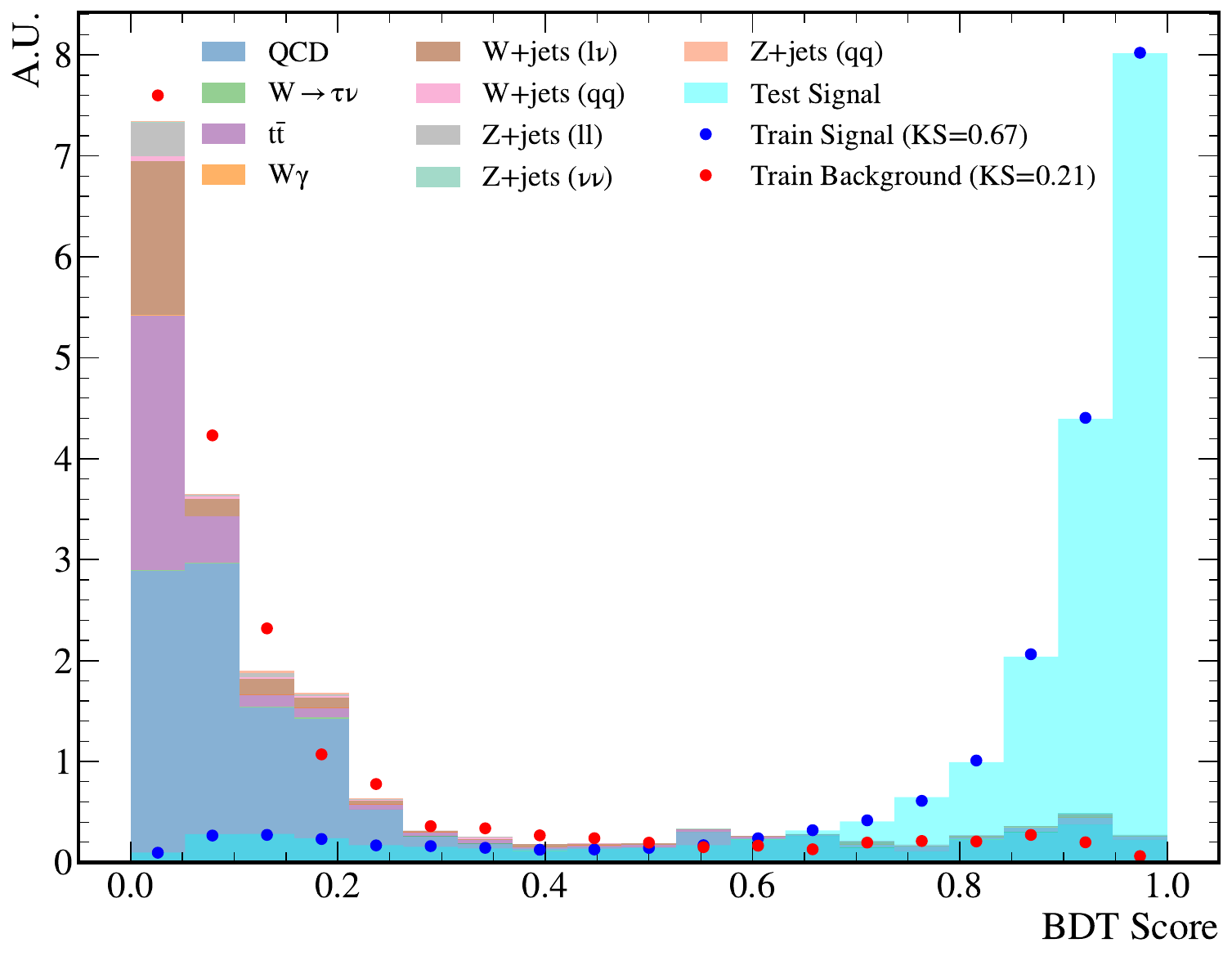}
    \caption{Normalized BDT score distributions for the $\wtopigamma$ signal and the inclusive standard model background. The plot compares the outputs evaluated on the training dataset (points) and the testing dataset (histograms). The consistency between the training and testing distributions, quantitatively supported by the Kolmogorov-Smirnov test results, demonstrates the absence of overtraining in the XGBoost classifier.}
    \label{fig:kstest}
\end{figure}

The robustness of the BDT training is rigorously evaluated to rule out any potential overtraining. Kolmogorov-Smirnov (KS) tests are performed on the BDT output score distributions, comparing the training and testing datasets ~\cite{ParticleDataGroup:2026aaa}. As illustrated in Fig.~\ref{fig:kstest}, the BDT scores for both the signal and background processes exhibit excellent agreement between the training and testing samples. This confirms that the XGBoost classifier is well-behaved, highly generalizable, and ready for final signal extraction.

To justify the joint application of the event-level BDT and the jet-level $\dpia$ discriminant, it is essential to rigorously evaluate their mutual correlation. If these two discriminators were highly correlated, imposing a stringent threshold on the BDT score would unintentionally distort the background distributions of some feature variables. Such phase-space distortion would violate the fundamental assumption of shape invariance across different selection regions.

\begin{figure}[htbp]
    \centering
    \includegraphics[width=0.95\linewidth]{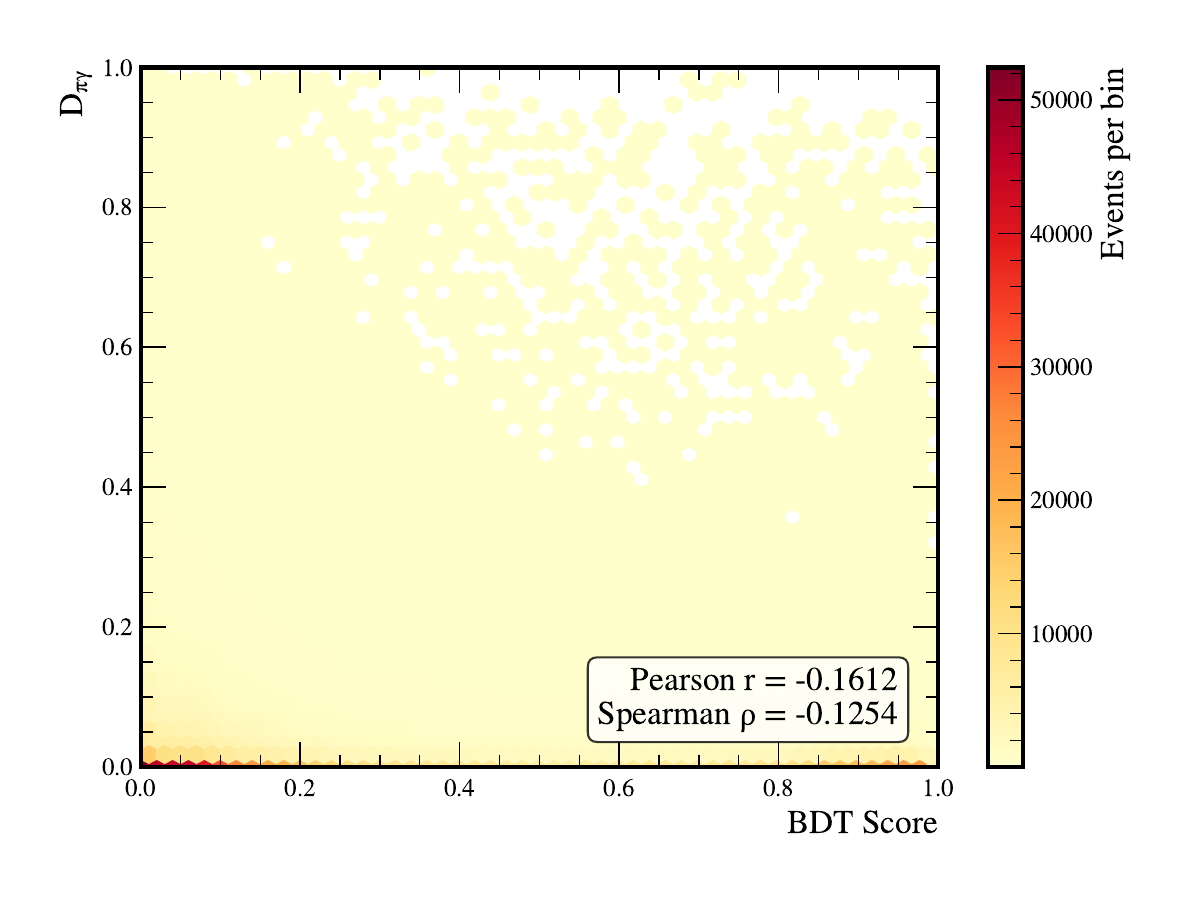}
    \caption{Two-dimensional distribution of the combined standard model backgrounds in the plane defined by the BDT score and the $\dpia$ discriminant. The near-zero correlation coefficients demonstrate the approximate independence between the event-level kinematics and the jet-level Sophon discriminant, supporting their combined use in the final signal extraction.}
    \label{fig:corre}
\end{figure}

To verify the independence of these two classifiers, we analyze the two-dimensional distribution of the combined Standard Model backgrounds in the plane of the BDT score and the Sophon discriminant $\dpia$, as presented in Fig.~\ref{fig:corre}. The Pearson correlation coefficient is  $r=-0.1612$ ~\cite{pearson1896vii}, with a corresponding Spearman's rank correlation coefficient of $\rho = -0.1254$ ~\cite{spearman1904proof}. These values indicate a very weak negative correlation, demonstrating that the macroscopic event kinematics captured by the BDT and the microscopic, internal jet-substructure features learned by Sophon remain largely decoupled.

This near-zero correlation ensures that applying a selection cut on the BDT score does not bias the background shape in either the $m_{\text{SD}}$ or other feature variables. Consequently, this independence provides a solid and unbiased foundation for the subsequent signal extraction.

\begin{figure}[htbp]
    \centering
    \includegraphics[width=0.95\linewidth]{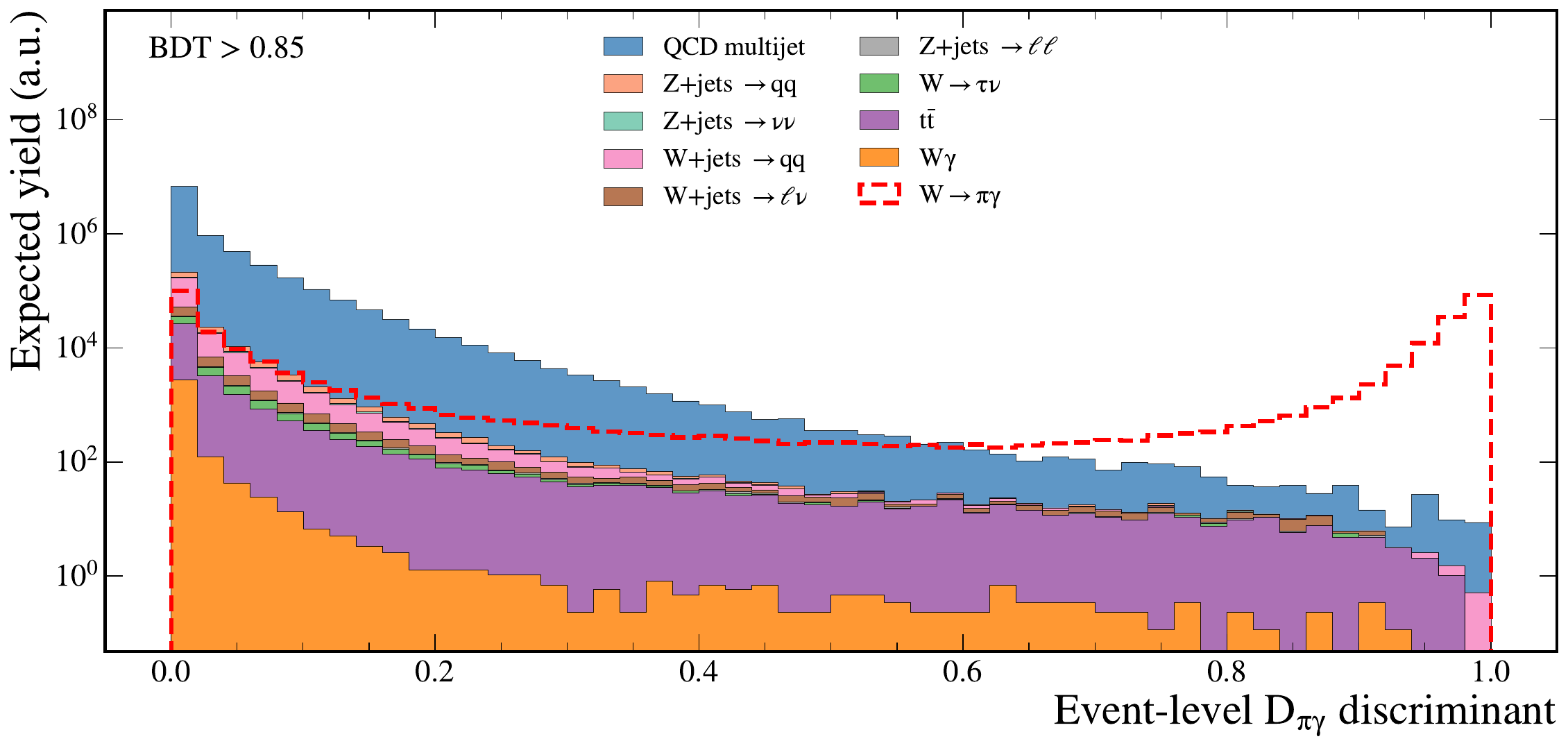}
    \caption{The stacked $\dpia$ distribution weighted by their respective physical cross-sections under $\mathrm{BDT}>0.85$. The pronounced signal peak near $\dpia \approx 1$ demonstrates the exceptional discrimination power and targeted background suppression achieved by incorporating the $P_1$ penalty topologies.  Here we set the hypothetical branching fraction ${\cal B}(\wtopigamma) = 1$.}
    \label{fig:SophonDpia}
\end{figure}

For the BDT score cut, we use a fixed BDT working point, $\mathrm{BDT}>0.85$. At this working point, most of the signal is preserved. Meanwhile, to ensure sufficient statistics for the subsequent background distribution extrapolation, more background is retained compared to a more stringent BDT cut, without excessively degrading the signal significance. Under the $\mathrm{BDT}>0.85$ requirement, the optimized discriminant $\dpia$ still possesses a powerful capability to distinguish the $\wtopigamma$ signal from the dominant Standard Model backgrounds. This impressive separation performance is clearly demonstrated in Fig.~\ref{fig:SophonDpia}, which presents the $\dpia$ distributions for both signal and background processes, normalized to an integrated luminosity of 450~$\text{fb}^{-1}$. The signal events are overwhelmingly clustered in the ultra-high score region near unity, whereas the massive QCD multijet background drops sharply by several orders of magnitude.

\section{Signal extraction}

\begin{figure}[htbp]
    \centering
    \includegraphics[width=0.95\linewidth]{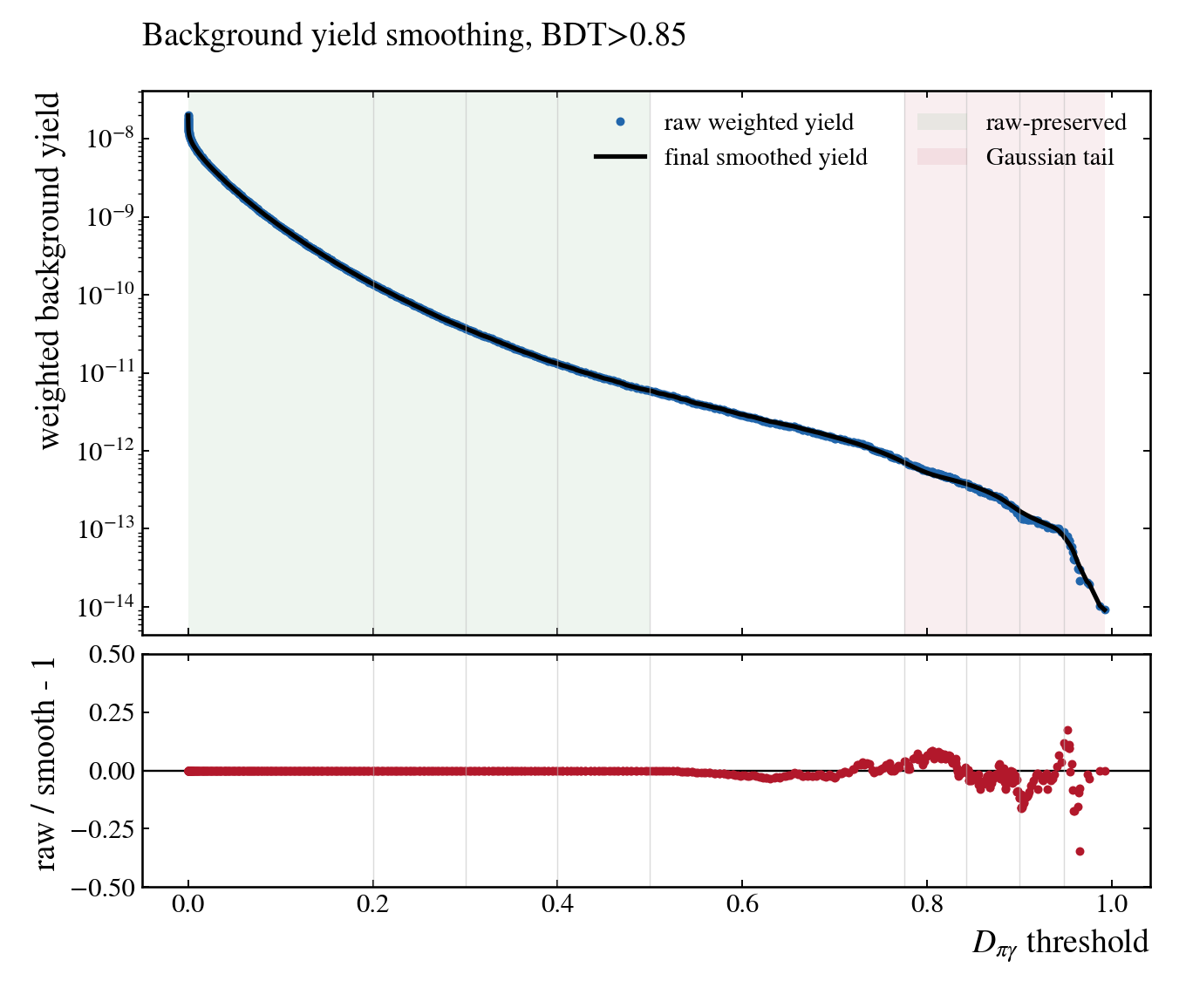}
    \caption{Background-yield smoothing as a function of the $\dpia$ threshold for the final $\mathrm{BDT}>0.85$ selection.  The raw weighted yield is preserved in the high-statistics low-$\dpia$ region, while the sparse high-$\dpia$ tail is smoothed in $x=-\ln(1-\dpia)$ and projected to be monotonic.}
    \label{fig:yieldSmoothing}
\end{figure}

\begin{figure*}[!t]
    \centering
    \includegraphics[width=0.98\textwidth]{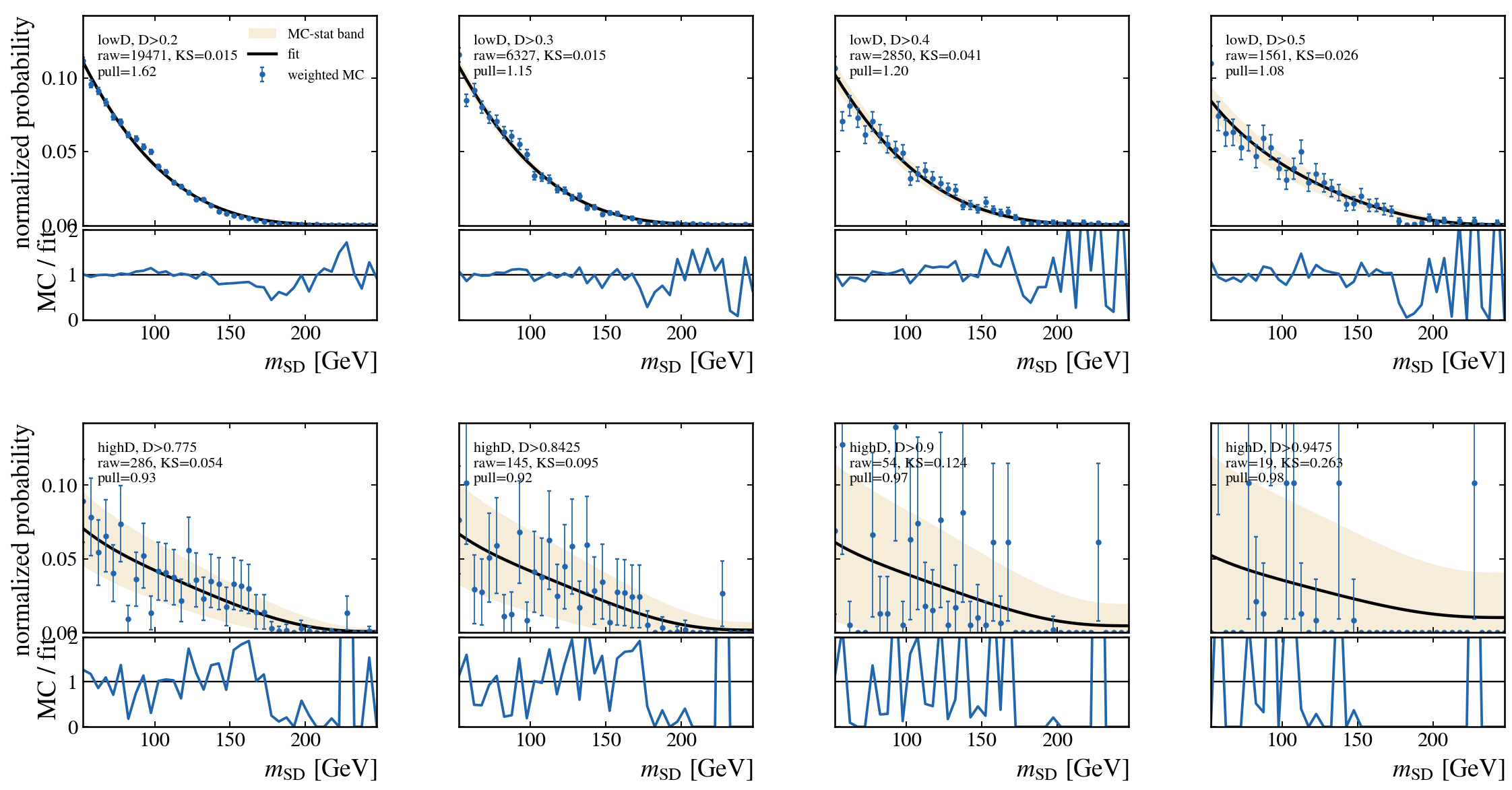}
    \caption{Validation of the final $m_{\mathrm{SD}}$ background shape model at representative $\dpia$ thresholds. Four low-$\dpia$ and four high-$\dpia$ points are shown.  In each panel, the weighted simulated background template is compared with the monotone Bernstein fit, and the lower ratio panel displays the agreement relative to the fitted shape.  The high-$\dpia$ region has limited simulated statistics, which motivates the smooth trajectory model and the additional uncertainty treatment used in the sensitivity estimate.}
    \label{fig:shapeFit}
\end{figure*}

Due to the extremely limited background statistics under high $\dpia$ score region, we implement background shape extrapolation for the final signal extraction. We scan thresholds on $\dpia$ under $\mathrm{BDT}>0.85$. Instead of relying on a single cut-and-count signal region, we construct a binned template in the soft-drop-mass range $50<m_{\mathrm{SD}}<250\, \gev$.  For each $\dpia$ threshold, the total background normalization is first obtained from the weighted simulated samples and then smoothed as a function of $x=-\ln(1-\dpia)$ using a Gaussian kernel, with the high-statistics low-$\dpia$ region preserved and a monotonic projection imposed, as shown in Fig.~\ref{fig:yieldSmoothing}.  The normalized $m_{\mathrm{SD}}$ shape is modeled by a monotone Bernstein polynomial of degree five,
\begin{equation}
    f(m_{\mathrm{SD}}|\dpia)=\sum_{k=0}^{5} c_k(\dpia) B_k(m_{\mathrm{SD}}),
\end{equation}
where the coefficients satisfy $c_0\ge c_1\ge\cdots\ge c_5\ge0$ through the parametrization $c_k=c_{k-1}\,\mathrm{sigmoid}(\theta_k)$.  The $\dpia$ dependence is captured by quadratic trajectories, $\theta_k(\dpia)=a_k\dpia^2+b_k\dpia+d_k$, fitted globally to weighted background templates with sufficient simulated statistics.  Fig.~\ref{fig:shapeFit} shows representative low- and high-$\dpia$ validation points for this smooth background model.

The expected sensitivity is evaluated from the binned signal and background templates as
\begin{equation}
    Z_{\mathrm{shape}}=\left[\sum_i \frac{N_{S,i}^2}{N_{B,i}+\sigma_{\mathrm{shape},i}^2}\right]^{1/2},
\end{equation}
where $N_{S,i}$ and $N_{B,i}$ are the expected signal and background yields in the $i$-th $m_{\mathrm{SD}}$ bin for an integrated luminosity of $450\,\mathrm{fb}^{-1}$~\cite{cowan2011asymptotic}.  The signal is normalized to a reference branching fraction ${\cal B}_{\rm ref}(\wtopigamma)=10^{-6}$.  The term $\sigma_{\mathrm{shape},i}$ is obtained from toy-data refits of the monotone Bernstein function and represents the shape-fit uncertainty.  The corresponding expected upper limit is estimated as ${\cal B}_{95}=1.64\,{\cal B}_{\rm ref}/Z_{\mathrm{shape}}$.

\begin{figure}[htbp]
    \centering
    \includegraphics[width=0.95\linewidth]{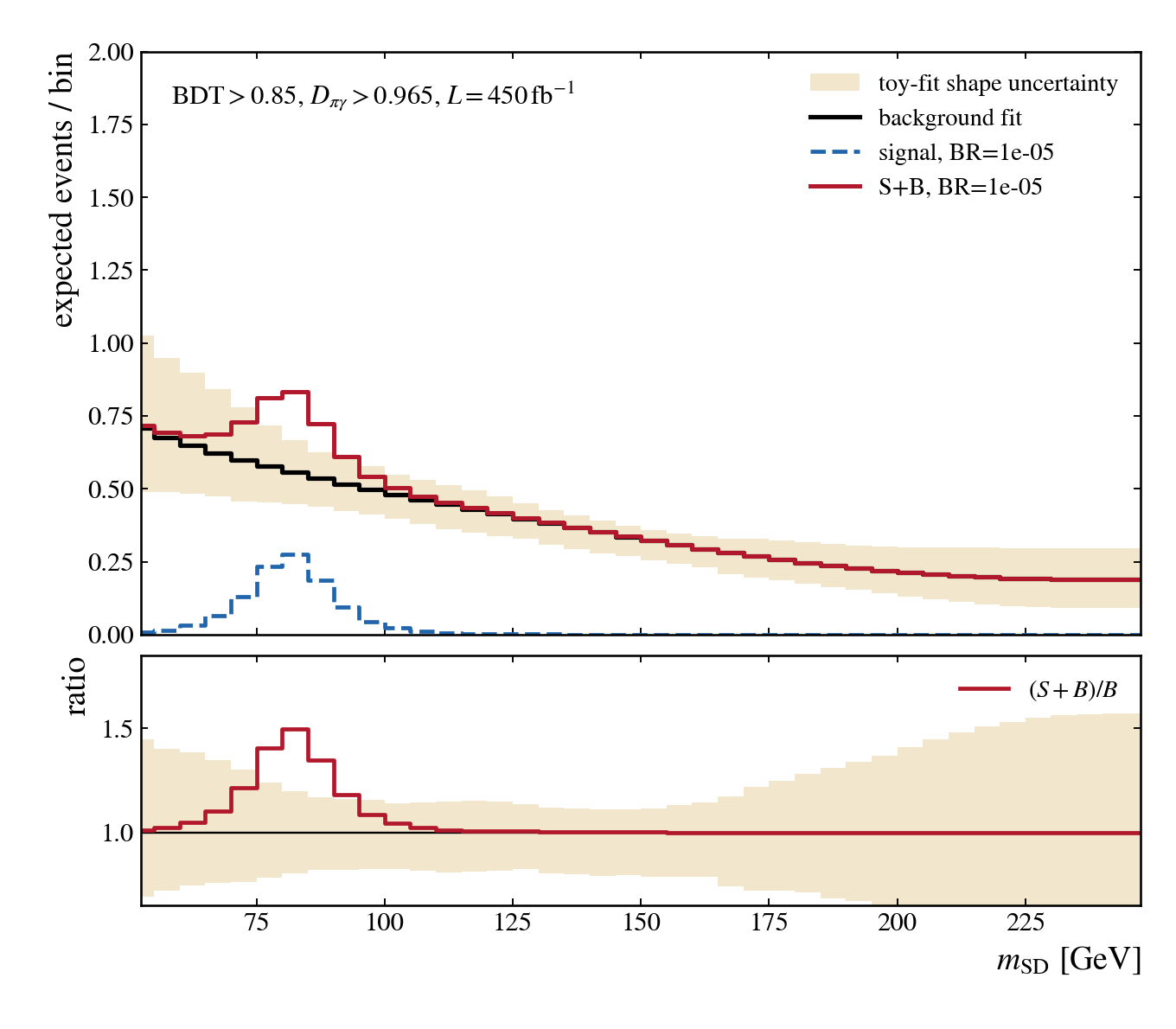}
    \caption{Final $m_{\mathrm{SD}}$ signal and background templates for the $N_B \ge10$ working point, $\mathrm{BDT}>0.85$ and $\dpia>0.965$.  The background prediction includes the toy-fit shape uncertainty band, while the signal overlay is shown with an enhanced normalization of ${\cal B}(\wtopigamma)=10^{-5}$ for visibility.}
    \label{fig:finalTemplate}
\end{figure}

Imposing a statistical threshold of at least ten expected background events, the best working point is found at $\mathrm{BDT}>0.85$ and $\dpia>0.965$.  Including the toy-fit shape uncertainty, this selection gives an expected upper limit of ${\cal B}_{95}(\wtopigamma)=2.78\times10^{-5}$ for $450\,\mathrm{fb}^{-1}$.  The corresponding signal and background templates are displayed in Fig.~\ref{fig:finalTemplate}.

\section{Conclusion}

In summary, we have presented a novel study on the ``hadron-in-fat-jet'' topology at the LHC, exploring a qualitatively new regime that connects jet substructure with exclusive hadronic states. Crucially, this work represents the first phenomenological sensitivity study of the rare decay $W^{\pm}\rightarrow\pi^{\pm}\gamma$ in the boosted regime, utilizing realistic fast detector simulation to model the complex signal and backgrounds. Unlike traditional substructure techniques targeting multi-prong partonic decays, our approach tackles hybrid configurations where a localized hadronic state is embedded within a single collimated jet.

To isolate this signature, we developed a comprehensive discrimination strategy that marks the first exploration of the Sophon model's fine-tuning methodology and its novel application to rare decay searches via jet tagging. By combining the fine-tuned, signature-oriented jet-level discriminant $D_{\pi\gamma}$ with an orthogonal event-level XGBoost BDT and a data-driven soft-drop-mass shape fit, we established an expected 95\% CL upper limit of $\mathcal{B}(W^{\pm}\rightarrow\pi^{\pm}\gamma) < 2.78\times10^{-5}$ for an integrated luminosity of $450\text{ fb}^{-1}$.

This framework enables further access to rare decay modes that are challenging in resolved topologies due to limited acceptance or overwhelming backgrounds, potentially enhancing sensitivity to both Standard Model measurements and new physics scenarios. Further improvements are expected by exploiting semileptonic $t\bar{t}$ events selected with single-lepton triggers, which provide a clean and efficiently triggered source of boosted $W$ bosons while substantially suppressing multijet backgrounds. The additional kinematic constraints from the top-quark system are anticipated to enhance signal purity and overall sensitivity. Beyond the specific $\wtopigamma$ benchmark, the proposed ``hadron-in-fat-jet'' strategy opens a new avenue for searching for light resonances decaying into a pion and a photon, whose highly boosted decay products may merge into a single jet and evade conventional reconstruction techniques. Our results demonstrate the potential of ``hadron-in-fat-jet'' tagging as a complementary paradigm to traditional boosted-object identification and establish a general framework for probing rare electroweak processes and light resonances at the LHC.

\FloatBarrier

\acknowledgments

This work is supported in part by the National Natural Science Foundation of China under Grants No.~12325504.

\newpage

\bibliography{WtoPiGamma}

\begin{thebibliography}{28}%
\makeatletter
\providecommand \@ifxundefined [1]{%
 \@ifx{#1\undefined}
}%
\providecommand \@ifnum [1]{%
 \ifnum #1\expandafter \@firstoftwo
 \else \expandafter \@secondoftwo
 \fi
}%
\providecommand \@ifx [1]{%
 \ifx #1\expandafter \@firstoftwo
 \else \expandafter \@secondoftwo
 \fi
}%
\providecommand \natexlab [1]{#1}%
\providecommand \enquote  [1]{``#1''}%
\providecommand \bibnamefont  [1]{#1}%
\providecommand \bibfnamefont [1]{#1}%
\providecommand \citenamefont [1]{#1}%
\providecommand \href@noop [0]{\@secondoftwo}%
\providecommand \href [0]{\begingroup \@sanitize@url \@href}%
\providecommand \@href[1]{\@@startlink{#1}\@@href}%
\providecommand \@@href[1]{\endgroup#1\@@endlink}%
\providecommand \@sanitize@url [0]{\catcode `\\12\catcode `\$12\catcode `\&12\catcode `\#12\catcode `\^12\catcode `\_12\catcode `\%12\relax}%
\providecommand \@@startlink[1]{}%
\providecommand \@@endlink[0]{}%
\providecommand \url  [0]{\begingroup\@sanitize@url \@url }%
\providecommand \@url [1]{\endgroup\@href {#1}{\urlprefix }}%
\providecommand \urlprefix  [0]{URL }%
\providecommand \Eprint [0]{\href }%
\providecommand \doibase [0]{https://doi.org/}%
\providecommand \selectlanguage [0]{\@gobble}%
\providecommand \bibinfo  [0]{\@secondoftwo}%
\providecommand \bibfield  [0]{\@secondoftwo}%
\providecommand \translation [1]{[#1]}%
\providecommand \BibitemOpen [0]{}%
\providecommand \bibitemStop [0]{}%
\providecommand \bibitemNoStop [0]{.\EOS\space}%
\providecommand \EOS [0]{\spacefactor3000\relax}%
\providecommand \BibitemShut  [1]{\csname bibitem#1\endcsname}%
\let\auto@bib@innerbib\@empty
\bibitem [{\citenamefont {Almeida}\ \emph {et~al.}(2009)\citenamefont {Almeida}, \citenamefont {Lee}, \citenamefont {Perez}, \citenamefont {Sterman}, \citenamefont {Sung},\ and\ \citenamefont {Virzi}}]{Almeida:2008yp}%
  \BibitemOpen
  \bibfield  {author} {\bibinfo {author} {\bibfnamefont {L.~G.}\ \bibnamefont {Almeida}}, \bibinfo {author} {\bibfnamefont {S.~J.}\ \bibnamefont {Lee}}, \bibinfo {author} {\bibfnamefont {G.}~\bibnamefont {Perez}}, \bibinfo {author} {\bibfnamefont {G.~F.}\ \bibnamefont {Sterman}}, \bibinfo {author} {\bibfnamefont {I.}~\bibnamefont {Sung}},\ and\ \bibinfo {author} {\bibfnamefont {J.}~\bibnamefont {Virzi}},\ }\bibfield  {title} {\bibinfo {title} {{Substructure of high-$p_T$ Jets at the LHC}},\ }\href {https://doi.org/10.1103/PhysRevD.79.074017} {\bibfield  {journal} {\bibinfo  {journal} {Phys. Rev. D}\ }\textbf {\bibinfo {volume} {79}},\ \bibinfo {pages} {074017} (\bibinfo {year} {2009})},\ \Eprint {https://arxiv.org/abs/0807.0234} {arXiv:0807.0234 [hep-ph]} \BibitemShut {NoStop}%
\bibitem [{\citenamefont {Marzani}\ \emph {et~al.}(2019)\citenamefont {Marzani}, \citenamefont {Soyez},\ and\ \citenamefont {Spannowsky}}]{Marzani:2019hun}%
  \BibitemOpen
  \bibfield  {author} {\bibinfo {author} {\bibfnamefont {S.}~\bibnamefont {Marzani}}, \bibinfo {author} {\bibfnamefont {G.}~\bibnamefont {Soyez}},\ and\ \bibinfo {author} {\bibfnamefont {M.}~\bibnamefont {Spannowsky}},\ }\href {https://doi.org/10.1007/978-3-030-15709-8} {\emph {\bibinfo {title} {{Looking Inside Jets: An Introduction to Jet Substructure and Boosted-object Phenomenology}}}},\ Vol.\ \bibinfo {volume} {958}\ (\bibinfo  {publisher} {Springer},\ \bibinfo {year} {2019})\ \Eprint {https://arxiv.org/abs/1901.10342} {arXiv:1901.10342 [hep-ph]} \BibitemShut {NoStop}%
\bibitem [{\citenamefont {Larkoski}\ \emph {et~al.}(2020)\citenamefont {Larkoski}, \citenamefont {Moult},\ and\ \citenamefont {Nachman}}]{Larkoski:2017jix}%
  \BibitemOpen
  \bibfield  {author} {\bibinfo {author} {\bibfnamefont {A.~J.}\ \bibnamefont {Larkoski}}, \bibinfo {author} {\bibfnamefont {I.}~\bibnamefont {Moult}},\ and\ \bibinfo {author} {\bibfnamefont {B.}~\bibnamefont {Nachman}},\ }\bibfield  {title} {\bibinfo {title} {{Jet Substructure at the Large Hadron Collider: A Review of Recent Advances in Theory and Machine Learning}},\ }\href {https://doi.org/10.1016/j.physrep.2019.11.001} {\bibfield  {journal} {\bibinfo  {journal} {Phys. Rept.}\ }\textbf {\bibinfo {volume} {841}},\ \bibinfo {pages} {1} (\bibinfo {year} {2020})},\ \Eprint {https://arxiv.org/abs/1709.04464} {arXiv:1709.04464 [hep-ph]} \BibitemShut {NoStop}%
\bibitem [{\citenamefont {Kogler}\ \emph {et~al.}(2019)\citenamefont {Kogler} \emph {et~al.}}]{Kogler:2018hem}%
  \BibitemOpen
  \bibfield  {author} {\bibinfo {author} {\bibfnamefont {R.}~\bibnamefont {Kogler}} \emph {et~al.},\ }\bibfield  {title} {\bibinfo {title} {{Jet Substructure at the Large Hadron Collider: Experimental Review}},\ }\href {https://doi.org/10.1103/RevModPhys.91.045003} {\bibfield  {journal} {\bibinfo  {journal} {Rev. Mod. Phys.}\ }\textbf {\bibinfo {volume} {91}},\ \bibinfo {pages} {045003} (\bibinfo {year} {2019})},\ \Eprint {https://arxiv.org/abs/1803.06991} {arXiv:1803.06991 [hep-ex]} \BibitemShut {NoStop}%
\bibitem [{\citenamefont {Grossman}\ \emph {et~al.}(2015)\citenamefont {Grossman}, \citenamefont {K{\"o}nig},\ and\ \citenamefont {Neubert}}]{grossman_exclusive_2015}%
  \BibitemOpen
  \bibfield  {author} {\bibinfo {author} {\bibfnamefont {Y.}~\bibnamefont {Grossman}}, \bibinfo {author} {\bibfnamefont {M.}~\bibnamefont {K{\"o}nig}},\ and\ \bibinfo {author} {\bibfnamefont {M.}~\bibnamefont {Neubert}},\ }\bibfield  {title} {\bibinfo {title} {{Exclusive Radiative Decays of W and Z Bosons in QCD Factorization}},\ }\href {https://doi.org/10.1007/JHEP04(2015)101} {\bibfield  {journal} {\bibinfo  {journal} {JHEP}\ }\textbf {\bibinfo {volume} {04}},\ \bibinfo {pages} {101}},\ \Eprint {https://arxiv.org/abs/1501.06569} {arXiv:1501.06569 [hep-ph]} \BibitemShut {NoStop}%
\bibitem [{\citenamefont {Melia}(2016)}]{melia_exclusive_2016}%
  \BibitemOpen
  \bibfield  {author} {\bibinfo {author} {\bibfnamefont {T.}~\bibnamefont {Melia}},\ }\bibfield  {title} {\bibinfo {title} {{Exclusive Hadronic $W$ Decay: $W \to \pi \gamma$ and $W \to \pi^+ \pi^+ \pi^-$}},\ }\href {https://doi.org/10.1016/j.nuclphysbps.2015.09.341} {\bibfield  {journal} {\bibinfo  {journal} {Nucl. Part. Phys. Proc.}\ }\textbf {\bibinfo {volume} {273}},\ \bibinfo {pages} {2102} (\bibinfo {year} {2016})}\BibitemShut {NoStop}%
\bibitem [{\citenamefont {Mangano}\ and\ \citenamefont {Melia}(2015)}]{mangano_rare_2015}%
  \BibitemOpen
  \bibfield  {author} {\bibinfo {author} {\bibfnamefont {M.}~\bibnamefont {Mangano}}\ and\ \bibinfo {author} {\bibfnamefont {T.}~\bibnamefont {Melia}},\ }\bibfield  {title} {\bibinfo {title} {{Rare Exclusive Hadronic $W$ Decays in a $t\bar{t}$ Environment}},\ }\href {https://doi.org/10.1140/epjc/s10052-015-3482-x} {\bibfield  {journal} {\bibinfo  {journal} {Eur. Phys. J. C}\ }\textbf {\bibinfo {volume} {75}},\ \bibinfo {pages} {258} (\bibinfo {year} {2015})},\ \Eprint {https://arxiv.org/abs/1410.7475} {arXiv:1410.7475 [hep-ph]} \BibitemShut {NoStop}%
\bibitem [{\citenamefont {Lepage}\ and\ \citenamefont {Brodsky}(1979)}]{PETERLEPAGE1979359}%
  \BibitemOpen
  \bibfield  {author} {\bibinfo {author} {\bibfnamefont {G.~P.}\ \bibnamefont {Lepage}}\ and\ \bibinfo {author} {\bibfnamefont {S.~J.}\ \bibnamefont {Brodsky}},\ }\bibfield  {title} {\bibinfo {title} {{Exclusive Processes in Quantum Chromodynamics: Evolution Equations for Hadronic Wave Functions and the Form-Factors of Mesons}},\ }\href {https://doi.org/10.1016/0370-2693(79)90554-9} {\bibfield  {journal} {\bibinfo  {journal} {Phys. Lett. B}\ }\textbf {\bibinfo {volume} {87}},\ \bibinfo {pages} {359} (\bibinfo {year} {1979})}\BibitemShut {NoStop}%
\bibitem [{\citenamefont {Aaltonen}\ \emph {et~al.}(2012)\citenamefont {Aaltonen} \emph {et~al.}}]{aaltonen_search_2012-1}%
  \BibitemOpen
  \bibfield  {author} {\bibinfo {author} {\bibfnamefont {T.}~\bibnamefont {Aaltonen}} \emph {et~al.} (\bibinfo {collaboration} {CDF}),\ }\bibfield  {title} {\bibinfo {title} {{Search for the Rare Radiative Decay: $W\rightarrow\pi\gamma$ in $p\bar{p}$ Collisions at $\sqrt{s} = 1.96$ TeV}},\ }\href {https://doi.org/10.1103/PhysRevD.85.032001} {\bibfield  {journal} {\bibinfo  {journal} {Phys. Rev. D}\ }\textbf {\bibinfo {volume} {85}},\ \bibinfo {pages} {032001} (\bibinfo {year} {2012})},\ \Eprint {https://arxiv.org/abs/1104.1585} {arXiv:1104.1585 [hep-ex]} \BibitemShut {NoStop}%
\bibitem [{\citenamefont {Sirunyan}\ \emph {et~al.}(2021)\citenamefont {Sirunyan} \emph {et~al.}}]{sirunyan_search_2021}%
  \BibitemOpen
  \bibfield  {author} {\bibinfo {author} {\bibfnamefont {A.~M.}\ \bibnamefont {Sirunyan}} \emph {et~al.} (\bibinfo {collaboration} {CMS}),\ }\bibfield  {title} {\bibinfo {title} {{Search for the Rare Decay of the W Boson into a Pion and a Photon in Proton-proton Collisions at $\sqrt{s}=13$ TeV}},\ }\href {https://doi.org/10.1016/j.physletb.2021.136409} {\bibfield  {journal} {\bibinfo  {journal} {Phys. Lett. B}\ }\textbf {\bibinfo {volume} {819}},\ \bibinfo {pages} {136409} (\bibinfo {year} {2021})},\ \Eprint {https://arxiv.org/abs/2011.06028} {arXiv:2011.06028 [hep-ex]} \BibitemShut {NoStop}%
\bibitem [{\citenamefont {Aad}\ \emph {et~al.}(2024)\citenamefont {Aad} \emph {et~al.}}]{collaboration_search_2024}%
  \BibitemOpen
  \bibfield  {author} {\bibinfo {author} {\bibfnamefont {G.}~\bibnamefont {Aad}} \emph {et~al.} (\bibinfo {collaboration} {ATLAS}),\ }\bibfield  {title} {\bibinfo {title} {{Search for the Exclusive W Boson Hadronic Decays $W^{\pm}\rightarrow \pi^{\pm}\gamma, W^{\pm}\rightarrow K^{\pm}\gamma$ and $W^{\pm}\rightarrow \rho^{\pm} \gamma$ with the ATLAS Detector}},\ }\href {https://doi.org/10.1103/PhysRevLett.133.161804} {\bibfield  {journal} {\bibinfo  {journal} {Phys. Rev. Lett.}\ }\textbf {\bibinfo {volume} {133}},\ \bibinfo {pages} {161804} (\bibinfo {year} {2024})},\ \Eprint {https://arxiv.org/abs/2309.15887} {arXiv:2309.15887 [hep-ex]} \BibitemShut {NoStop}%
\bibitem [{\citenamefont {Alwall}\ \emph {et~al.}(2014)\citenamefont {Alwall}, \citenamefont {Frederix}, \citenamefont {Frixione}, \citenamefont {Hirschi}, \citenamefont {Maltoni}, \citenamefont {Mattelaer}, \citenamefont {Shao}, \citenamefont {Stelzer}, \citenamefont {Torrielli},\ and\ \citenamefont {Zaro}}]{alwall2014automated}%
  \BibitemOpen
  \bibfield  {author} {\bibinfo {author} {\bibfnamefont {J.}~\bibnamefont {Alwall}}, \bibinfo {author} {\bibfnamefont {R.}~\bibnamefont {Frederix}}, \bibinfo {author} {\bibfnamefont {S.}~\bibnamefont {Frixione}}, \bibinfo {author} {\bibfnamefont {V.}~\bibnamefont {Hirschi}}, \bibinfo {author} {\bibfnamefont {F.}~\bibnamefont {Maltoni}}, \bibinfo {author} {\bibfnamefont {O.}~\bibnamefont {Mattelaer}}, \bibinfo {author} {\bibfnamefont {H.-S.}\ \bibnamefont {Shao}}, \bibinfo {author} {\bibfnamefont {T.}~\bibnamefont {Stelzer}}, \bibinfo {author} {\bibfnamefont {P.}~\bibnamefont {Torrielli}},\ and\ \bibinfo {author} {\bibfnamefont {M.}~\bibnamefont {Zaro}},\ }\bibfield  {title} {\bibinfo {title} {{The Automated Computation of Tree-level and Next-to-leading Order Differential Cross Sections, and Their Matching to Parton Shower Simulations}},\ }\href {https://doi.org/10.1007/JHEP07(2014)079} {\bibfield  {journal} {\bibinfo  {journal} {JHEP}\ }\textbf {\bibinfo {volume} {07}}},\ \Eprint {https://arxiv.org/abs/1405.0301} {arXiv:1405.0301 [hep-ph]} \BibitemShut {NoStop}%
\bibitem [{\citenamefont {Ball}\ \emph {et~al.}(2017)\citenamefont {Ball} \emph {et~al.}}]{ball2017parton}%
  \BibitemOpen
  \bibfield  {author} {\bibinfo {author} {\bibfnamefont {R.~D.}\ \bibnamefont {Ball}} \emph {et~al.},\ }\bibfield  {title} {\bibinfo {title} {{Parton Distributions from High-precision Collider Data}},\ }\href {https://doi.org/10.1140/epjc/s10052-017-5199-5} {\bibfield  {journal} {\bibinfo  {journal} {Eur. Phys. J. C}\ }\textbf {\bibinfo {volume} {77}},\ \bibinfo {pages} {663} (\bibinfo {year} {2017})},\ \Eprint {https://arxiv.org/abs/1706.00428} {arXiv:1706.00428 [hep-ph]} \BibitemShut {NoStop}%
\bibitem [{\citenamefont {Sjostrand}\ \emph {et~al.}(2008)\citenamefont {Sjostrand}, \citenamefont {Mrenna},\ and\ \citenamefont {Skands}}]{Sjostrand:2007gs}%
  \BibitemOpen
  \bibfield  {author} {\bibinfo {author} {\bibfnamefont {T.}~\bibnamefont {Sjostrand}}, \bibinfo {author} {\bibfnamefont {S.}~\bibnamefont {Mrenna}},\ and\ \bibinfo {author} {\bibfnamefont {P.~Z.}\ \bibnamefont {Skands}},\ }\bibfield  {title} {\bibinfo {title} {{A Brief Introduction to PYTHIA 8.1}},\ }\href {https://doi.org/10.1016/j.cpc.2008.01.036} {\bibfield  {journal} {\bibinfo  {journal} {Comput. Phys. Commun.}\ }\textbf {\bibinfo {volume} {178}},\ \bibinfo {pages} {852} (\bibinfo {year} {2008})},\ \Eprint {https://arxiv.org/abs/0710.3820} {arXiv:0710.3820 [hep-ph]} \BibitemShut {NoStop}%
\bibitem [{\citenamefont {de~Favereau}\ \emph {et~al.}(2014)\citenamefont {de~Favereau}, \citenamefont {Delaere}, \citenamefont {Demin}, \citenamefont {Giammanco}, \citenamefont {Lema{\^\i}tre}, \citenamefont {Mertens},\ and\ \citenamefont {Selvaggi}}]{deFavereau:2013fsa}%
  \BibitemOpen
  \bibfield  {author} {\bibinfo {author} {\bibfnamefont {J.}~\bibnamefont {de~Favereau}}, \bibinfo {author} {\bibfnamefont {C.}~\bibnamefont {Delaere}}, \bibinfo {author} {\bibfnamefont {P.}~\bibnamefont {Demin}}, \bibinfo {author} {\bibfnamefont {A.}~\bibnamefont {Giammanco}}, \bibinfo {author} {\bibfnamefont {V.}~\bibnamefont {Lema{\^\i}tre}}, \bibinfo {author} {\bibfnamefont {A.}~\bibnamefont {Mertens}},\ and\ \bibinfo {author} {\bibfnamefont {M.}~\bibnamefont {Selvaggi}},\ }\bibfield  {title} {\bibinfo {title} {{DELPHES 3, A Modular Framework for Fast Simulation of a Generic Collider Experiment}},\ }\href {https://doi.org/10.1007/JHEP02(2014)057} {\bibfield  {journal} {\bibinfo  {journal} {JHEP}\ }\textbf {\bibinfo {volume} {02}},\ \bibinfo {pages} {1}},\ \Eprint {https://arxiv.org/abs/1307.6346} {arXiv:1307.6346 [hep-ex]} \BibitemShut {NoStop}%
\bibitem [{\citenamefont {Li}\ \emph {et~al.}(2024)\citenamefont {Li} \emph {et~al.}}]{li_accelerating_2024}%
  \BibitemOpen
  \bibfield  {author} {\bibinfo {author} {\bibfnamefont {C.}~\bibnamefont {Li}} \emph {et~al.},\ }\bibfield  {title} {\bibinfo {title} {{Accelerating Resonance Searches via Signature-Oriented Pre-training}},\ }\href {https://doi.org/10.1038/s42005-024-01887-6} {\bibfield  {journal} {\bibinfo  {journal} {Commun. Phys.}\ }\textbf {\bibinfo {volume} {7}},\ \bibinfo {pages} {405} (\bibinfo {year} {2024})},\ \Eprint {https://arxiv.org/abs/2405.12972} {arXiv:2405.12972 [hep-ph]} \BibitemShut {NoStop}%
\bibitem [{\citenamefont {Bertolini}\ \emph {et~al.}(2014)\citenamefont {Bertolini}, \citenamefont {Harris}, \citenamefont {Low},\ and\ \citenamefont {Tran}}]{bertolini2014pileup}%
  \BibitemOpen
  \bibfield  {author} {\bibinfo {author} {\bibfnamefont {D.}~\bibnamefont {Bertolini}}, \bibinfo {author} {\bibfnamefont {P.}~\bibnamefont {Harris}}, \bibinfo {author} {\bibfnamefont {M.}~\bibnamefont {Low}},\ and\ \bibinfo {author} {\bibfnamefont {N.}~\bibnamefont {Tran}},\ }\bibfield  {title} {\bibinfo {title} {{Pileup Per Particle Identification}},\ }\href {https://doi.org/10.1007/JHEP10(2014)059} {\bibfield  {journal} {\bibinfo  {journal} {JHEP}\ }\textbf {\bibinfo {volume} {10}},\ \bibinfo {pages} {1}},\ \Eprint {https://arxiv.org/abs/1407.6013} {arXiv:1407.6013 [hep-ph]} \BibitemShut {NoStop}%
\bibitem [{\citenamefont {Hayrapetyan}\ \emph {et~al.}(2024)\citenamefont {Hayrapetyan}, \citenamefont {Tumasyan}, \citenamefont {Adam}, \citenamefont {Andrejkovic}, \citenamefont {Benato}, \citenamefont {Bergauer}, \citenamefont {Chatterjee}, \citenamefont {Damanakis}, \citenamefont {Dragicevic}, \citenamefont {Hussain} \emph {et~al.}}]{hayrapetyan2024performance}%
  \BibitemOpen
  \bibfield  {author} {\bibinfo {author} {\bibfnamefont {A.}~\bibnamefont {Hayrapetyan}}, \bibinfo {author} {\bibfnamefont {A.}~\bibnamefont {Tumasyan}}, \bibinfo {author} {\bibfnamefont {W.}~\bibnamefont {Adam}}, \bibinfo {author} {\bibfnamefont {J.}~\bibnamefont {Andrejkovic}}, \bibinfo {author} {\bibfnamefont {L.}~\bibnamefont {Benato}}, \bibinfo {author} {\bibfnamefont {T.}~\bibnamefont {Bergauer}}, \bibinfo {author} {\bibfnamefont {S.}~\bibnamefont {Chatterjee}}, \bibinfo {author} {\bibfnamefont {K.}~\bibnamefont {Damanakis}}, \bibinfo {author} {\bibfnamefont {M.}~\bibnamefont {Dragicevic}}, \bibinfo {author} {\bibfnamefont {P.}~\bibnamefont {Hussain}}, \emph {et~al.},\ }\bibfield  {title} {\bibinfo {title} {Performance of the cms high-level trigger during lhc run 2},\ }\href@noop {} {\bibfield  {journal} {\bibinfo  {journal} {Journal of Instrumentation}\ }\textbf {\bibinfo {volume} {19}}\bibinfo  {number} { (11)},\ \bibinfo {pages} {P11021}}\BibitemShut {NoStop}%
\bibitem [{\citenamefont {Zhao}\ \emph {et~al.}(2025)\citenamefont {Zhao}, \citenamefont {Li}, \citenamefont {Agapitos}, \citenamefont {Fu}, \citenamefont {Gao}, \citenamefont {Mao},\ and\ \citenamefont {Li}}]{zhao_novel_2025}%
  \BibitemOpen
\bibfield  {number} {  }\bibfield  {author} {\bibinfo {author} {\bibfnamefont {Y.}~\bibnamefont {Zhao}}, \bibinfo {author} {\bibfnamefont {C.}~\bibnamefont {Li}}, \bibinfo {author} {\bibfnamefont {A.}~\bibnamefont {Agapitos}}, \bibinfo {author} {\bibfnamefont {D.}~\bibnamefont {Fu}}, \bibinfo {author} {\bibfnamefont {L.}~\bibnamefont {Gao}}, \bibinfo {author} {\bibfnamefont {Y.}~\bibnamefont {Mao}},\ and\ \bibinfo {author} {\bibfnamefont {Q.}~\bibnamefont {Li}},\ }\href@noop {} {\bibinfo {title} {{Novel $|V_{cb}|$ Extraction Method via Boosted $bc$-tagging with In-situ Calibration}}} (\bibinfo {year} {2025}),\ \Eprint {https://arxiv.org/abs/2503.00118} {arXiv:2503.00118 [hep-ph]} \BibitemShut {NoStop}%
\bibitem [{\citenamefont {collaboration}\ \emph {et~al.}(2020)\citenamefont {collaboration} \emph {et~al.}}]{cms2020identification}%
  \BibitemOpen
  \bibfield  {author} {\bibinfo {author} {\bibfnamefont {C.}~\bibnamefont {collaboration}} \emph {et~al.},\ }\bibfield  {title} {\bibinfo {title} {Identification of heavy, energetic, hadronically decaying particles using machine-learning techniques},\ }\href@noop {} {\bibfield  {journal} {\bibinfo  {journal} {arXiv preprint arXiv:2004.08262}\ } (\bibinfo {year} {2020})}\BibitemShut {NoStop}%
\bibitem [{\citenamefont {Qu}\ and\ \citenamefont {Gouskos}(2020)}]{qu2020jet}%
  \BibitemOpen
  \bibfield  {author} {\bibinfo {author} {\bibfnamefont {H.}~\bibnamefont {Qu}}\ and\ \bibinfo {author} {\bibfnamefont {L.}~\bibnamefont {Gouskos}},\ }\bibfield  {title} {\bibinfo {title} {Jet tagging via particle clouds},\ }\href@noop {} {\bibfield  {journal} {\bibinfo  {journal} {Physical Review D}\ }\textbf {\bibinfo {volume} {101}},\ \bibinfo {pages} {056019} (\bibinfo {year} {2020})}\BibitemShut {NoStop}%
\bibitem [{\citenamefont {Dasgupta}\ \emph {et~al.}(2013)\citenamefont {Dasgupta}, \citenamefont {Fregoso}, \citenamefont {Marzani},\ and\ \citenamefont {Salam}}]{dasgupta2013towards}%
  \BibitemOpen
  \bibfield  {author} {\bibinfo {author} {\bibfnamefont {M.}~\bibnamefont {Dasgupta}}, \bibinfo {author} {\bibfnamefont {A.}~\bibnamefont {Fregoso}}, \bibinfo {author} {\bibfnamefont {S.}~\bibnamefont {Marzani}},\ and\ \bibinfo {author} {\bibfnamefont {G.~P.}\ \bibnamefont {Salam}},\ }\bibfield  {title} {\bibinfo {title} {{Towards an Understanding of Jet Substructure}},\ }\href {https://doi.org/10.1007/JHEP09(2013)029} {\bibfield  {journal} {\bibinfo  {journal} {JHEP}\ }\textbf {\bibinfo {volume} {2013}}\bibfield  {number} {\bibinfo  {number} { (9)},\ \bibinfo {pages} {1}},\ }\Eprint {https://arxiv.org/abs/1307.0007} {arXiv:1307.0007 [hep-ph]} \BibitemShut {NoStop}%
\bibitem [{\citenamefont {Larkoski}\ \emph {et~al.}(2014)\citenamefont {Larkoski}, \citenamefont {Marzani}, \citenamefont {Soyez},\ and\ \citenamefont {Thaler}}]{larkoski2014soft}%
  \BibitemOpen
  \bibfield  {author} {\bibinfo {author} {\bibfnamefont {A.~J.}\ \bibnamefont {Larkoski}}, \bibinfo {author} {\bibfnamefont {S.}~\bibnamefont {Marzani}}, \bibinfo {author} {\bibfnamefont {G.}~\bibnamefont {Soyez}},\ and\ \bibinfo {author} {\bibfnamefont {J.}~\bibnamefont {Thaler}},\ }\bibfield  {title} {\bibinfo {title} {{Soft Drop}},\ }\href {https://doi.org/10.1007/JHEP05(2014)146} {\bibfield  {journal} {\bibinfo  {journal} {JHEP}\ }\textbf {\bibinfo {volume} {05}},\ \bibinfo {pages} {1}},\ \Eprint {https://arxiv.org/abs/1402.2657} {arXiv:1402.2657 [hep-ph]} \BibitemShut {NoStop}%
\bibitem [{\citenamefont {Chen}\ and\ \citenamefont {Guestrin}(2016)}]{chen2016xgboost}%
  \BibitemOpen
  \bibfield  {author} {\bibinfo {author} {\bibfnamefont {T.}~\bibnamefont {Chen}}\ and\ \bibinfo {author} {\bibfnamefont {C.}~\bibnamefont {Guestrin}},\ }\bibfield  {title} {\bibinfo {title} {Xgboost: A scalable tree boosting system},\ }in\ \href@noop {} {\emph {\bibinfo {booktitle} {Proceedings of the 22nd ACM SIGKDD International Conference on Knowledge Discovery and Data Mining}}}\ (\bibinfo {year} {2016})\ pp.\ \bibinfo {pages} {785--794}\BibitemShut {NoStop}%
\bibitem [{\citenamefont {Takahashi}\ \emph {et~al.}(2026)\citenamefont {Takahashi} \emph {et~al.}}]{ParticleDataGroup:2026aaa}%
  \BibitemOpen
  \bibfield  {author} {\bibinfo {author} {\bibfnamefont {F.}~\bibnamefont {Takahashi}} \emph {et~al.} (\bibinfo {collaboration} {Particle Data Group}),\ }\bibfield  {title} {\bibinfo {title} {{Review of Particle Physics}},\ }\href {https://doi.org/10.1142/S0217751X26300115} {\bibfield  {journal} {\bibinfo  {journal} {Int. J. Mod. Phys. A}\ }\textbf {\bibinfo {volume} {41}},\ \bibinfo {pages} {2630011} (\bibinfo {year} {2026})}\BibitemShut {NoStop}%
\bibitem [{\citenamefont {Pearson}(1896)}]{pearson1896vii}%
  \BibitemOpen
  \bibfield  {author} {\bibinfo {author} {\bibfnamefont {K.}~\bibnamefont {Pearson}},\ }\bibfield  {title} {\bibinfo {title} {Vii. mathematical contributions to the theory of evolution.—iii. regression, heredity, and panmixia},\ }\href@noop {} {\bibfield  {journal} {\bibinfo  {journal} {Philosophical Transactions of the Royal Society of London. Series A, containing papers of a mathematical or physical character}\ ,\ \bibinfo {pages} {253}} (\bibinfo {year} {1896})}\BibitemShut {NoStop}%
\bibitem [{\citenamefont {Spearman}(1904)}]{spearman1904proof}%
  \BibitemOpen
  \bibfield  {author} {\bibinfo {author} {\bibfnamefont {C.}~\bibnamefont {Spearman}},\ }\bibfield  {title} {\bibinfo {title} {The proof and measurement of association between two things},\ }\href@noop {} {\bibfield  {journal} {\bibinfo  {journal} {The American Journal of Psychology}\ }\textbf {\bibinfo {volume} {15}},\ \bibinfo {pages} {72} (\bibinfo {year} {1904})}\BibitemShut {NoStop}%
\bibitem [{\citenamefont {Cowan}\ \emph {et~al.}(2011)\citenamefont {Cowan}, \citenamefont {Cranmer}, \citenamefont {Gross},\ and\ \citenamefont {Vitells}}]{cowan2011asymptotic}%
  \BibitemOpen
  \bibfield  {author} {\bibinfo {author} {\bibfnamefont {G.}~\bibnamefont {Cowan}}, \bibinfo {author} {\bibfnamefont {K.}~\bibnamefont {Cranmer}}, \bibinfo {author} {\bibfnamefont {E.}~\bibnamefont {Gross}},\ and\ \bibinfo {author} {\bibfnamefont {O.}~\bibnamefont {Vitells}},\ }\bibfield  {title} {\bibinfo {title} {Asymptotic formulae for likelihood-based tests of new physics},\ }\href@noop {} {\bibfield  {journal} {\bibinfo  {journal} {The European Physical Journal C}\ }\textbf {\bibinfo {volume} {71}},\ \bibinfo {pages} {1554} (\bibinfo {year} {2011})}\BibitemShut {NoStop}%
\end{thebibliography}%

\end{document}